\documentclass{article}
\usepackage[preprint]{spconf}
\usepackage{amsmath,graphicx}
\usepackage[hidelinks]{hyperref}
\usepackage{url}
\usepackage{multirow}
\usepackage{wrapfig}
\usepackage{booktabs}
\usepackage{listings}
\usepackage{enumitem}
\usepackage{rotating}
\usepackage{makecell}
\usepackage{caption} 
\usepackage{algorithm}
\usepackage{enumitem}
\usepackage{algpseudocode}
\usepackage{amsthm,mathtools}
\usepackage{highlight}
\usepackage{colortbl}
\toappear{To appear in {\it  2024 IEEE Spoken Language Technology Workshop, Dec 02-05, 2024, Macao, China}}
\title{Privacy versus Emotion Preservation Trade-offs in Emotion-Preserving Speaker Anonymization}
\name{\it{Zexin Cai, Henry Li Xinyuan, Ashi Garg, Leibny Paola Garc\'ia-Perera, Kevin Duh,} \\ \it Sanjeev Khudanpur, Nicholas Andrews, Matthew Wiesner}
\address{Human Language Technology Center of Excellence, Johns Hopkins University}

\begin{document}
\ninept
\maketitle
\begin{abstract}
Advances in speech technology now allow unprecedented access to personally identifiable information through speech. To protect such information, the differential privacy field has explored ways to anonymize speech while preserving its utility, including linguistic and paralinguistic aspects. However, anonymizing speech while maintaining emotional state remains challenging. We explore this problem in the context of the VoicePrivacy 2024 challenge. Specifically, we developed various speaker anonymization pipelines and find that approaches either excel at anonymization or preserving emotion state, but not both simultaneously. Achieving both would require an in-domain emotion recognizer. Additionally, we found that it is feasible to train a semi-effective speaker verification system using only emotion representations, demonstrating the challenge of separating these two modalities.
\end{abstract}
\section{Introduction}
\label{sec:intro}
Voice-driven interaction has been integrated into various aspects of human life, making tasks more convenient and hands-free. This technology has seen significant growth in the modern era, with notable examples including virtual assistants on smart devices, wearable technology, and customer service applications.
However, the increasing use of voice-driven interaction raises security and privacy concerns, particularly regarding the exposure of speech recordings to fraudsters and hackers when transmitted over untrusted public networks~\cite{nautsch2019preserving}. Consequently, the personally identifiable information in the raw speech signal can be susceptible to leakage or extraction~\cite{zhang2023voicepm}. 

To mitigate privacy concerns associated with the potential interception and misuse of speech data, speech anonymization is employed to protect the most sensitive information, speaker identity, within speech. Specifically, speech anonymization aims to suppress acoustic characteristics that could be used to identify the speaker while at the same time preserving other characteristics, chiefly linguistic content, within the speech. The field of speech anonymization is still nascent, with formal definitions and a comparison platform for solutions on standardized datasets and protocols recently established by the VoicePrivacy Challenge series~\cite{tomashenko2020introducing}.

Since speech anonymization inherently involves altering and transforming speech, most research has centered on techniques such as voice conversion (VC), speech synthesis, noise addition, and traditional signal processing methods to achieve anonymization~\cite{tomashenko2020introducing}. Among the developed anonymization techniques, the x-vector-based method~\cite{fang19_ssw}, used as the baseline for the VoicePrivacy challenge, offers a flexible choice of pseudo-speaker and achieves adequate performance in privacy and utility assessments. Essentially, the x-vector-based method employs a framework similar to an any-to-any VC approach, synthesizing anonymized speech by conditioning the framework with x-vector~\cite{snyder2018xvector} speaker representations to produce pseudo-speakers' voices. Several subsequent studies have improved the x-vector-based method from various angles to boost its privacy protection ability~\cite{miao2023speaker}, such as constructing x-vectors via singular value modification~\cite{mawalim20_interspeech} and using a generative model to sample pseudo-speakers in the x-vector space~\cite{turner2020speaker}. Beyond the approaches described above that achieve speech anonymization through acoustic models like those used in VC techniques, other research explores a speech synthesis-based method by cascading automatic speech recognition (ASR) and text-to-speech (TTS) systems~\cite{meyer2023prosody, meyer22b_interspeech}, which can significantly eliminate speaker identity footprints in speech. 

Recent developments in the speech anonymization community have presented a more complex anonymization scenario. Besides preserving linguistic content and hiding speaker identity, an anonymization system should also maintain unchanged paralinguistic attributes~\cite{tomashenko2022voiceprivacy}. Under these conditions, researchers struggle to conceal speaker identity while retaining paralinguistic attributes, highlighting the trade-off between utility (paralinguistic attributes) and privacy (speaker identification) in this setting~\cite{zhang2023voicepm}. The VoicePrivacy Challenge 2024 emphasizes the preservation of emotional state~\cite{tomashenko2024voiceprivacy}. Additionally, the challenge recognizes the risk that an attacker could access anonymized data and train a new speaker verification model on it. Therefore, understanding how emotion and speaker information are entangled in speech signals is essential in this anonymization context to overcome the privacy-utility trade-off.

\begin{figure*}[ht]
  \centering
  \includegraphics[width=0.76\textwidth]{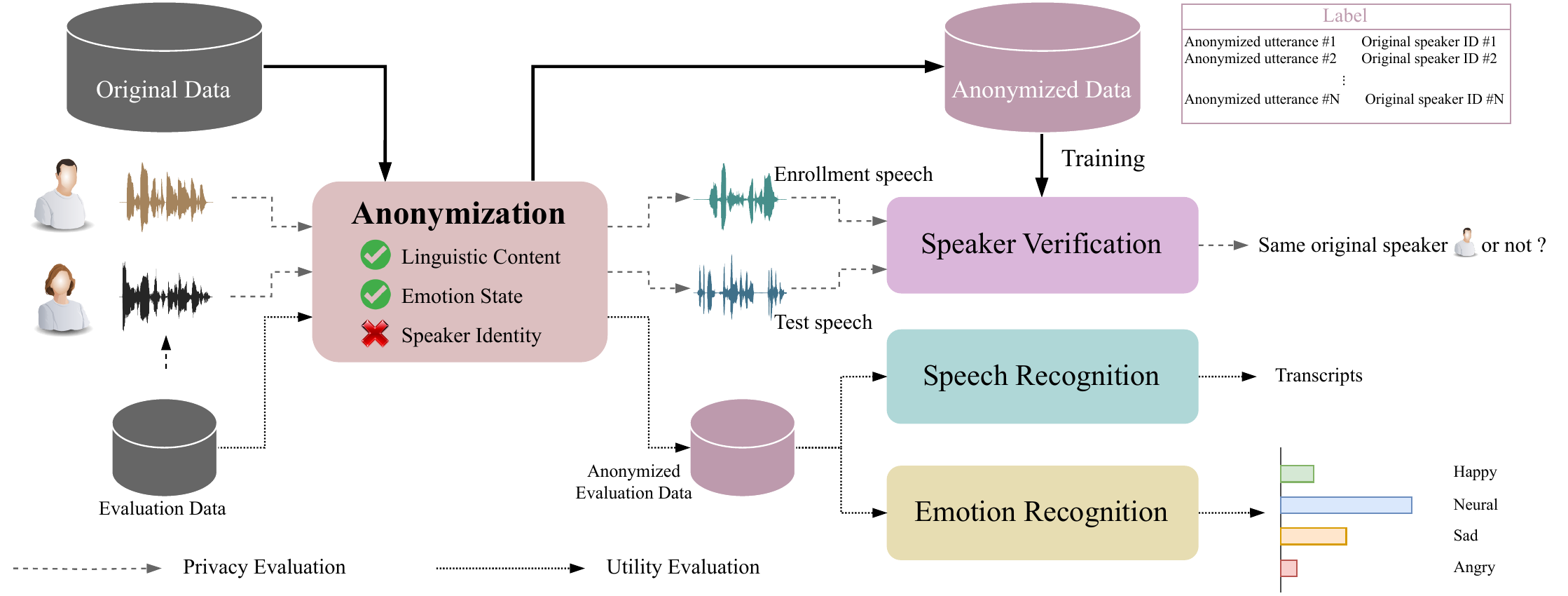}
  \caption{Speech anonymization task and evaluation pipeline (w.r.t the VoicePrivacy 2024 Challenge)}
  \label{fig:anon_process}
\end{figure*}

Earlier work on anonymization has explored the privacy-utility trade-off, but does not investigate its causes or potential solutions~\cite{zhang2023voicepm}. Aside from voice timbre, prosodic features such as melody, rhythm, and intensity—shaped by a speaker’s social environment and critical learning period—also provide significant information about their identity~\cite{mary2006prosodic}. This theory is supported by findings that the source speaker can be recognized to a certain degree in voice-converted speech~\cite{cai2023identifying}. Inspired by the above research, this paper delves into the factors that might cause the leakage of speaker identity and investigates the relationship between speaker and emotion in speech. To achieve this, we apply various VC-based and cascaded ASR-TTS methods to the anonymization task in the VoicePrivacy Challenge 2024. Our study reveals that speech emotion recognition (SER) and automatic speaker verification (ASV) systems rely on overlapping speech attributes. Disentangling identity from acoustic properties is a non-trivial task, as these properties are closely related. 
While we can minimize the trade-off between privacy and emotion preservation given prior knowledge of the corresponding in-domain emotion recognizer, the challenge of separating speaker and emotion information in speech remains significant. Finally, our results suggest that emotion recognizers can serve as a reliable objective evaluation metric in emotional speech synthesis.

\section{Method}
\subsection{Task Definition and Evaluation Metrics}
In the semi-informed speech anonymization task a user supplies speech data and attempts to protect their identity using a speech anonymization system. An attacker, who has access to the anonymized data, attempts to discover the speaker's identity. A speech anonymization system is created to obscure the user’s identity while maintaining the linguistic content and paralinguistic attributes. In the VoicePrivacy Challenge 2024, the goal is to maintain the emotional state of the speech post-anonymization. As such, the anonymization performance is evaluated from two fronts: privacy evaluation and utility evaluation. The privacy metric measures how well the system conceals the original speakers’ identities, whereas the utility metrics access the retention of content and emotional state.

Figure~\ref{fig:anon_process} illustrates the anonymization and evaluation pipeline of this task. The core element is the anonymization module, which converts each original input audio into anonymized audio under the following conditions: 1. preserving linguistic content, 2. preserving emotional state, and 3. removing speaker identity information.

The privacy evaluation pipeline adheres to a standard speaker verification process. The verification model is trained on anonymized data labeled with the original speakers’ identities. An effective anonymization system should sufficiently distort and obscure the original identities at the waveform level, preventing the speaker verification system from identifying different speakers. During evaluation, pairs of source speech from the evaluation dataset are anonymized and treated as enrollment and test speech. The speaker verification model then assesses whether the two utterances originate from the same original speaker. With a perfect anonymization system, the verification system, acting as the attacker,  performs no better than random guessing. The main metric for privacy evaluation is the equal error rate (EER), calculated based on similarity scores from pairs of utterances in the anonymized evaluation set, known as trials. A lower EER indicates a greater risk of speaker re-identification,  thus a higher EER indicates better performance in preserving voice privacy.


For utility evaluation, the anonymized evaluation data is transcribed using a speech recognition system. The performance in preserving content is assessed by comparing these transcripts to the ground truth content from the source data and measuring the word error rate (WER). Similarly, an emotion recognizer is employed on the anonymized data to determine the emotional state of the anonymized speech. In this case, four emotion states—Happy, Neutral, Sad, and Angry—are evaluated. An anonymization system demonstrates good emotion preservation performance if the emotional state of the anonymized speech matches that of the original speech. Preservation of emotion state is measured using the Unweighted average recall (UAR)~\cite{tomashenko2024voiceprivacy}. In general, A lower WER denotes superior preservation of linguistic content, whereas a higher UAR indicates superior preservation of emotion states.


\subsection{Anonymization Approaches}
\label{sec:anon_app}
\begin{figure}[ht]
    \centering
    \includegraphics[width=0.45\textwidth]{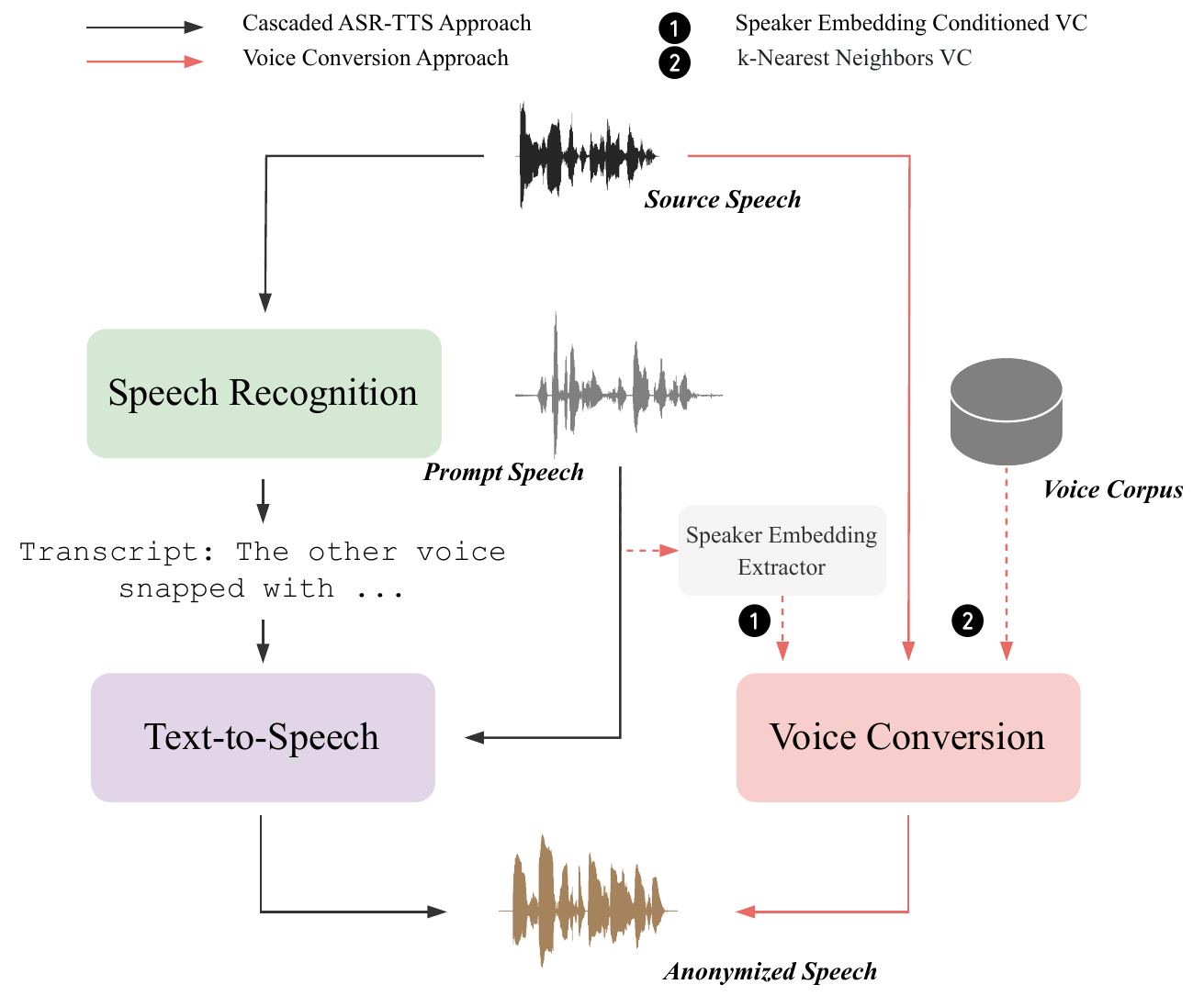}
    \caption{VC-based and cascaded ASR-TTS anonymization process}
    \label{fig:anonproc}
\end{figure}

We employ two primary synthesis approaches to achieve anonymization for the described task. The speech anonymization process is shown in Figure~\ref{fig:anonproc}, where one method is based on voice conversion (VC) models, and the other employs a cascaded ASR-TTS pipeline.

VC is a method that changes the voice of the source speech to match that of a target speaker, preserving the content and most prosodic features. This technique aligns closely with the objectives of the anonymization task. We explore two VC-based systems that can convert source utterances to a variety of target speakers.

The first VC-based anonymization system, similar to the x-vector-based system from the VoicePrivacy Challenge, conditions on speaker representations to convert the source voice to a target speaker’s voice. The model uses the content representation extracted by a pre-trained self-supervised learning (SSL) model called ContentVec~\cite{qian2022ContentVec} as input. This approach uses a transformer-based VC system~\cite{CAI2023cross} to convert the input content representation into the target speaker’s Mel-spectrogram by conditioning on the target speaker’s representation vector. The audio waveform is then reconstructed from the Mel-spectrogram using a HiFi-GAN vocoder~\cite{hifigan}.

Another VC-based solution utilizes kNN VC~\cite{knn_vc}, functioning at the WavLM-feature~\cite{wavlm} level. The kNN-VC system maps the WavLM features of the source utterance to those of the target speaker using k-nearest neighbor regression. Each frame from the source speech is replaced by the average of the k-nearest neighbor target WavLM features, followed by a HiFi-GAN vocoder to synthesize the target utterance. This approach, unlike the previous VC method, necessitates a target speech corpus for conversion.

While VC-based systems can effectively modify the acoustic characteristics related to the timbre of source speakers, certain prosodic features, reflecting the speakers’ habitual speaking styles, remain unchanged. These prosodic features could be used to identify the speaker. To mitigate this, we employ a cascaded ASR-TTS pipeline to enhance anonymization by modifying the speaking style of the source speech. As shown in Figure~\ref{fig:anonproc}, we first transcribe the source utterance using an ASR system. Subsequently, a multi-speaker TTS system generates the anonymized utterance, cloning the voice and speaking style from a prompt utterance.

\section{Experiments}
In this section, we explore the relationship between emotion and speaker information using the anonymization pipeline from the VoicePrivacy 2024 challenge, focusing on English corpora. We anonymized datasets using the systems in Section \ref{sec:anon_app} and evaluated their privacy and utility performance. For the cascaded approach, we used various utterance prompts, from randomized speech to audio containing some original speaking style, to study the impact on speaker identity exposure. After identifying the trade-off between privacy and emotion preservation, we explored strategies to mitigate it. Additionally, we analyzed the extraction of speaker information from emotion embeddings to understand their overlap and the challenges in disentangling them.

\subsection{Dataset}
The VoicePrivacy 2024 challenge uses subsets from the LibriSpeech~\cite{librispeech} and IEMOCAP~\cite{busso2008iemocap} corpora for development and evaluation. More details can be found in the data description section of the challenge’s evaluation plan~\cite{tomashenko2024voiceprivacy}. There are 10 subsets specifically designated for the evaluation process. The subsets libri-dev-asr and libri-test-asr are used for ASR evaluation. The subsets libri-dev-enrolls, libri-dev-trials-f, libri-dev-trials-m, libri-test-enrolls, libri-test-trials-f, and libri-test-trials-m are used for evaluating privacy (speaker verification) performance. The subset libri-train-clean-360 is employed for training the speaker verification system following anonymization. For emotion preservation performance, the subsets IEMOCAP-dev and IEMOCAP-test are utilized.

We also incorporate the LibriTTS~\cite{zen2019libritts} speech synthesis dataset in our experiments. This dataset comprises 585 hours of clean speech data at 24kHz from 2,456 speakers. We also utilize the VoxCeleb1~\cite{nagrani17_interspeech} dataset in our study, with the training data including 148,642 utterances from 1,211 speakers and the test set comprising 4,874 utterances from 40 speakers. There is no overlap between the training and testing data in both the source and target datasets.

\subsection{Experimental Details}
We train the speaker embedding-conditioned VC model outlined in Section~\ref{sec:anon_app} using the LibriTTS training sets. Content features are extracted with the pre-trained ContentVec\_legacy-500 model,\footnote{\scriptsize\url{https://github.com/auspicious3000/contentvec}} and speaker embeddings are obtained from an ECAPA-TDNN model~\cite{Desplanques2020ECAPATDNNEC} by SpeechBrain~\cite{speechbrain}. The VC conversion system, including the feature transformation and vocoder modules, is trained on audio recordings at a 24kHz sample rate. After anonymization, we downsample the synthesized audio to 16kHz for evaluation. For the kNN-VC method, $k$ is set to 4.

In the cascaded ASR-TTS approach, we employ the ‘medium-en’ model from Whisper\footnote{\scriptsize\url{https://github.com/openai/whisper}}~\cite{radford2023robust} as our ASR system to transcribe the source utterance. The Whisper model achieves a WER of 3.38\% on the libri-dev-asr set and 3.29\% on the libri-test-asr set. For the study of privacy and emotion preservation, we choose the open-source synthesis model XTTS,\footnote{\scriptsize\url{https://github.com/coqui-ai/TTS}} which is a generative TTS model providing high-fidelity synthesis and capable of voice and style cloning based on a prompt audio segment.

\subsection{Anonymization Performance}
We anonymized the utterances from the datasets selected by the VoicePrivacy challenge\footnote{\scriptsize\url{https://github.com/Voice-Privacy-Challenge/Voice-Privacy-Challenge-2024}} using various approaches detailed in Section~\ref{sec:anon_app}. 
The anonymization performance of different systems is summarized in Table~\ref{tab:knn_vc}, with the corresponding systems annotated as follows:

    
\begin{table*}[th]
  \caption{Privacy and utility performance of various anonymization approaches \\ (darker color indicates better performance)}
  \label{tab:knn_vc}
  \scriptsize
  \renewcommand{\arraystretch}{1.1}
  \centering
  \setlength{\tabcolsep}{3pt}
  \begin{tabular}{l ccccc ccc ccc}
    \toprule
    \multirow{2}{*}{\textbf{System}} & \multicolumn{5}{c}{\textbf{Privacy - EER} (\%) $\uparrow$} & \multicolumn{3}{c}{\textbf{Utility - UAR mean} (\%) $\uparrow$} & \multicolumn{3}{c}{\textbf{Utility - WER} (\%) $\downarrow$} \\ 
    \cmidrule(lr){2-6} \cmidrule(lr){7-9} \cmidrule(lr){10-12} 
    & libri-dev-f & libri-dev-m & libri-test-f & libri-test-m & avg. & IEMOCAP-dev & IEMOCAP-test & avg. & libri-dev & libri-test & avg. \\
    \midrule
    Origin & \gradientcell{10.511}{0}{50}{white}{violet}{30} & \gradientcell{0.931}{0}{50}{white}{violet}{30}  & \gradientcell{8.761}{0}{30}{white}{violet}{30}  & \gradientcell{0.418}{0}{30}{white}{violet}{30}  & \gradientcell{5.16}{0}{30}{white}{violet}{30}  & \gradientcell{69.0796}{0}{75}{white}{olive}{30}   & \gradientcell{71.0618}{0}{75}{white}{olive}{30}  & \gradientcell{70.07}{0}{75}{white}{olive}{30}  & \gradientcell{1.807}{0}{5}{teal}{white}{50}  & \gradientcell{1.844}{0}{5}{teal}{white}{50} & \gradientcell{1.83}{0}{5}{teal}{white}{50} \\
    ConVec2Mel-VC$^1$ & \gradientcell{15.342}{0}{50}{white}{violet}{30} & \gradientcell{7.451}{0}{50}{white}{violet}{30} & \gradientcell{10.444}{0}{50}{white}{violet}{30} & \gradientcell{5.57}{0}{50}{white}{violet}{30} & \gradientcell{9.70}{0}{50}{white}{violet}{30} & \gradientcell{50.7706}{0}{75}{white}{olive}{30}& \gradientcell{48.3282}{0}{75}{white}{olive}{30} & \gradientcell{49.55}{0}{75}{white}{olive}{30} & \gradientcell{2.157}{0}{5}{teal}{white}{50} & \gradientcell{2.269}{0}{5}{teal}{white}{50} &
    \gradientcell{2.21}{0}{5}{teal}{white}{50}\\
    kNN-VC$^1$ & \gradientcell{18.351}{0}{50}{white}{violet}{30} & \gradientcell{13.663}{0}{50}{white}{violet}{30} & \gradientcell{16.239}{0}{50}{white}{violet}{30} & \gradientcell{12.496}{0}{50}{white}{violet}{30} & \gradientcell{15.19}{0}{50}{white}{violet}{30} & \gradientcell{47.7042}{0}{75}{white}{olive}{30}& \gradientcell{50.6086}{0}{75}{white}{olive}{30} & \gradientcell{49.16}{0}{75}{white}{olive}{30} & \gradientcell{2.991}{0}{5}{teal}{white}{50} & \gradientcell{2.962}{0}{5}{teal}{white}{50} &
    \gradientcell{2.98}{0}{5}{teal}{white}{50}\\
    ConVec2Mel-VC-XTTS$^2$ & \gradientcell{39.775}{0}{50}{white}{violet}{30} & \gradientcell{31.056}{0}{50}{white}{violet}{30} & \gradientcell{36.817}{0}{50}{white}{violet}{30} & \gradientcell{30.959}{0}{50}{white}{violet}{30} & \gradientcell{34.65}{0}{50}{white}{violet}{30} & \gradientcell{45.2988}{0}{75}{white}{olive}{30}& \gradientcell{40.6318}{0}{75}{white}{olive}{30} & \gradientcell{42.97}{0}{75}{white}{olive}{30} & \gradientcell{3.999}{0}{5}{teal}{white}{50} & \gradientcell{4.329}{0}{5}{teal}{white}{50} &
    \gradientcell{4.16}{0}{5}{teal}{white}{50}\\
    kNN-VC-XTTS$^2$ & \gradientcell{44.034}{0}{50}{white}{violet}{30} & \gradientcell{44.567}{0}{50}{white}{violet}{30} & \gradientcell{43.939}{0}{50}{white}{violet}{30} & \gradientcell{46.135}{0}{50}{white}{violet}{30} & \gradientcell{44.67}{0}{50}{white}{violet}{30} & \gradientcell{36.7774}{0}{75}{white}{olive}{30}& \gradientcell{38.0922}{0}{75}{white}{olive}{30} & \gradientcell{37.43}{0}{75}{white}{olive}{30} & \gradientcell{4.758}{0}{5}{teal}{white}{50} & \gradientcell{4.069}{0}{5}{teal}{white}{50} & \gradientcell{4.41}{0}{5}{teal}{white}{50}     \\
    XTTS$^3$ & \gradientcell{48.143}{0}{50}{white}{violet}{30} & \gradientcell{48.769}{0}{50}{white}{violet}{30} & \gradientcell{47.040}{0}{50}{white}{violet}{30} & \gradientcell{47.660}{0}{50}{white}{violet}{30} & \gradientcell{47.90}{0}{50}{white}{violet}{30} & \gradientcell{34.3710}{0}{75}{white}{olive}{30}& \gradientcell{32.9232}{0}{75}{white}{olive}{30} & \gradientcell{33.65}{0}{75}{white}{olive}{30} & \gradientcell{4.869}{0}{5}{teal}{white}{50} & \gradientcell{4.537}{0}{5}{teal}{white}{50} &
    \gradientcell{4.70}{0}{5}{teal}{white}{50}\\
    \midrule
    Emo$_{\textit{MSP}}$-XTTS$^4$ & \gradientcell{44.034}{0}{50}{white}{violet}{30} & \gradientcell{37.888}{0}{50}{white}{violet}{30} & \gradientcell{46.899}{0}{50}{white}{violet}{30} & \gradientcell{45.637}{0}{50}{white}{violet}{30} & \gradientcell{43.61}{0}{50}{white}{violet}{30} & \gradientcell{36.8728}{0}{75}{white}{olive}{30}& \gradientcell{37.0036}{0}{75}{white}{olive}{30} & \gradientcell{36.94}{0}{75}{white}{olive}{30} & \gradientcell{4.834}{0}{5}{teal}{white}{50} & \gradientcell{3.898}{0}{5}{teal}{white}{50} & \gradientcell{4.37}{0}{5}{teal}{white}{50} \\
    Emo$_{\textit{IEMOCAP}}$-XTTS$^4$ & \gradientcell{43.751}{0}{50}{white}{violet}{30} & \gradientcell{44.100}{0}{50}{white}{violet}{30} & \gradientcell{45.256}{0}{50}{white}{violet}{30} & \gradientcell{47.834}{0}{50}{white}{violet}{30} & \gradientcell{45.24}{0}{50}{white}{violet}{30} & \gradientcell{52.0652}{0}{75}{white}{olive}{30}& \gradientcell{52.8012}{0}{75}{white}{olive}{30} & \gradientcell{52.43}{0}{75}{white}{olive}{30} & \gradientcell{4.520}{0}{5}{teal}{white}{50} & \gradientcell{3.967}{0}{5}{teal}{white}{50} & \gradientcell{4.24}{0}{5}{teal}{white}{50} \\
    \bottomrule
    \multicolumn{10}{l}{ \textit{$^1$alter the original voice while leaving some prosodic features, such as phoneme durations, unchanged} } \\
    \multicolumn{10}{l}{ \textit{$^2$clone the anonymized voice and speaking style from the anonymized speech} } \\
    \multicolumn{10}{l}{ \textit{$^3$fully anonymize the voice by cloning both the voice and speaking style from a random utterance} } \\
    \multicolumn{10}{l}{ \textit{$^4$replicate a different speaker’s voice and speaking style, yet expressing the same emotion} } \\
  \end{tabular}
\end{table*}

\begin{itemize}[noitemsep, leftmargin=*]
    \item Origin: The original, unanonymized speech.
    \item ConVec2Mel-VC: The speaker embedding-conditioned VC system we developed. During anonymization, the target embedding is extracted from a randomly selected utterance from LibriTTS.
    \item kNN-VC: The kNN-based VC method. For each utterance, the WavLM feature pool is obtained from a randomly chosen target speaker in LibriTTS, with the target pool comprising at least 5 minutes of audio.
    \item ConVec2Mel-VC-XTTS: This system utilizes the cascaded ASR-TTS method. For each source utterance, the prompt speech is the corresponding anonymized speech from ConVec2Mel-VC.
    \item kNN-VC-XTTS: This system follows the cascaded ASR-TTS approach. For each source utterance, the prompt speech is the corresponding anonymized speech from the kNN-VC system.
    \item XTTS: The cascaded ASR-TTS anonymization method, where the prompt utterance during inference is randomly chosen from the LibriTTS dataset.
\end{itemize}

As indicated in the table, VC-based anonymization systems perform better in emotion preservation. Both ConVec2Mel-VC and kNN-VC show comparable performance, with an average UAR of around 49\%, implying that the speech attributes retained by these systems support the emotion recognizer in identifying the target emotion. Nevertheless, some hidden speech characteristics tied to speaker identity remain unanonymized, allowing the speaker verification model to detect these patterns, resulting in an average EER of less than 20\%. Specifically, ConVec2Mel-VC achieves an EER of 9.7\% in privacy evaluation, while the kNN-VC approach attains an average EER of around 15.19\%. Therefore, although the VC anonymization systems preserve some level of emotion, they also leak speaker information. Both systems maintain the original content well, as reflected in the WER results.


The XTTS system achieves the highest privacy performance among all systems, with an EER close to 50\%, by cloning a random voice and speaking style from another utterance. However, when the XTTS system is conditioned on an utterance with a modified voice but preserved prosodic attributes, resulting anonymization systems like ConVec2Mel-VC-XTTS and kNN-VC-XTTS can achieve higher emotion preservation scores. For instance, the ConVec2Mel-VC-XTTS system, which clones the speaking style from the ConVec2Mel-VC system, achieves an average UAR of 42.97\%. This is lower than the UAR of ConVec2Mel-VC but higher than the XTTS approach, which has a UAR of 33.65\%. Notably, the emotion preservation performance of kNN-VC-XTTS, while better than XTTS, is not significantly higher. This might be due to the lack of temporal coherence in the kNN-VC, leading to a distorted distribution compared to normal speech and causing the XTTS system to struggle with cloning the corresponding speaking style. 

This again leads to speaker identity leakage from these preserved attributes. Regarding privacy preservation, the ConVec2Mel-VC-XTTS system achieves an EER of about 34.65\%, which is lower than the XTTS approach. This suggests that attributes other than voice timbre can reveal speaker information and are helpful in emotion recognition.

\subsection{Achieving the best of both worlds}
\begin{figure}[ht]
    \centering
    \includegraphics[width=0.34\textwidth]{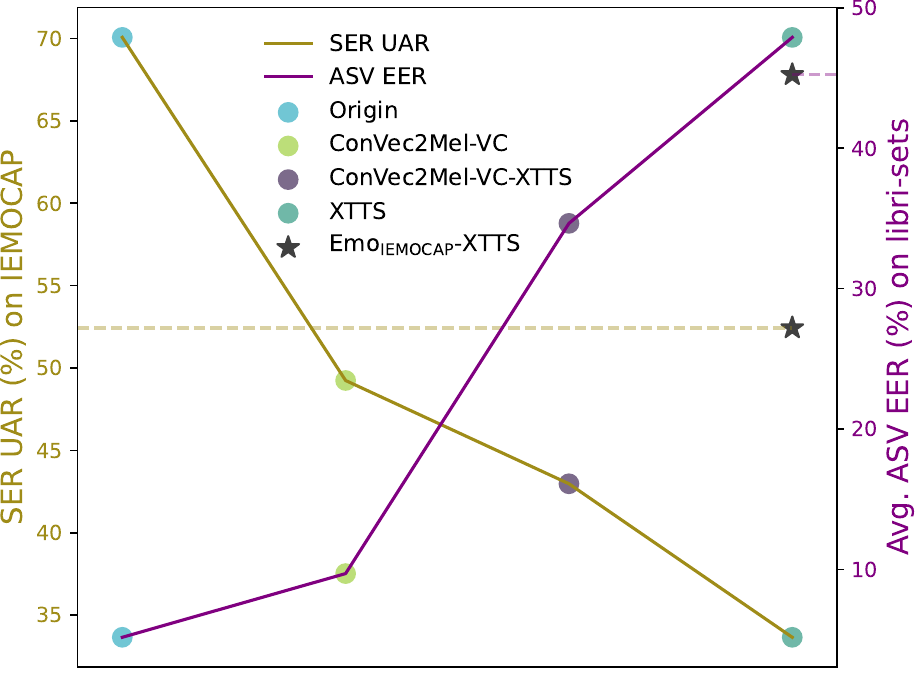}
    \caption{Privacy-emotion preservation trade-off}
    \label{fig:tradeoff}
\end{figure}

The systems discussed above demonstrate a clear trade-off between privacy and emotion preservation performance. The results are shown in Figure \ref{fig:tradeoff}. As emotion preservation performance rises, speaker information leakage takes place, leading to a decrease in privacy performance.

Based on the above results, we propose that randomly cloning a speaker’s voice with a different speaking style expressing the same emotion could break this trade-off. To test this, we use emotion embeddings extracted from emotion recognizers as a proxy to find a target utterance for voice and style cloning in the cascaded ASR-TTS approach. This study examines two types of emotion embeddings. The first is a concatenated emotion representation from embeddings extracted by five pre-trained emotion recognizers provided by the challenge. This in-domain representation, derived from systems trained with IEMOCAP, represents the optimal emotion preservation achievable when the anonymization system has access to the SER system. The second embedding is obtained from an out-of-domain extractor\footnote{\scriptsize\url{https://huggingface.co/audeering/wav2vec2-large-robust-12-ft-emotion-msp-dim}}~\cite{10089511} trained with the MSP-Podcast dataset~\cite{Lotfian_2019_3}.

For each utterance to be anonymized, the prompt audio is selected from the LibriTTS dataset using the following steps: 1. randomly select 5000 utterances from the dataset, 2. calculate the cosine similarity between the emotion representations of source utterance and all 5000 utterances, 3. randomly choose one target utterance from the top 10 utterances with the highest similarity scores.

The last two lines of Table \ref{tab:knn_vc} present the results of the anonymization system that employs the emotion-proxy anonymization strategy. Emo$_{\textit{MSP}}$-XTTS is based on the out-of-domain emotion recognizer, while Emo$_{\textit{IEMOCAP}}$-XTTS relies on the in-domain emotion recognizer. Emo$_{\textit{IEMOCAP}}$-XTTS achieves strong emotion preservation performance with a UAR of 52.43\% and, simultaneously, high privacy performance with an EER of 45.24\%. This system is marked in Figure \ref{fig:tradeoff} as the ideal system, breaking the privacy-emotion preservation trade-off shown earlier. However, this assumes that the anonymization system has prior knowledge of the emotion recognition system. 

In the alternative scenario, where the anonymization system possesses out-of-domain prior emotion knowledge, Emo$_{\textit{MSP}}$-XTTS achieves privacy performance comparable to Emo$_{\textit{IEMOCAP}}$-XTTS but struggles to find the best match for emotion, with an emotion preservation UAR of 36.94\%. Despite this, the emotion preservation score is higher than that of the XTTS system, suggesting that the emotion-proxy strategy is effective in the anonymization task.

\subsection{Speaker-Identifying Information in Emotion Embeddings}
\begin{figure}[ht]
    \centering
\includegraphics[width=0.335\textwidth]{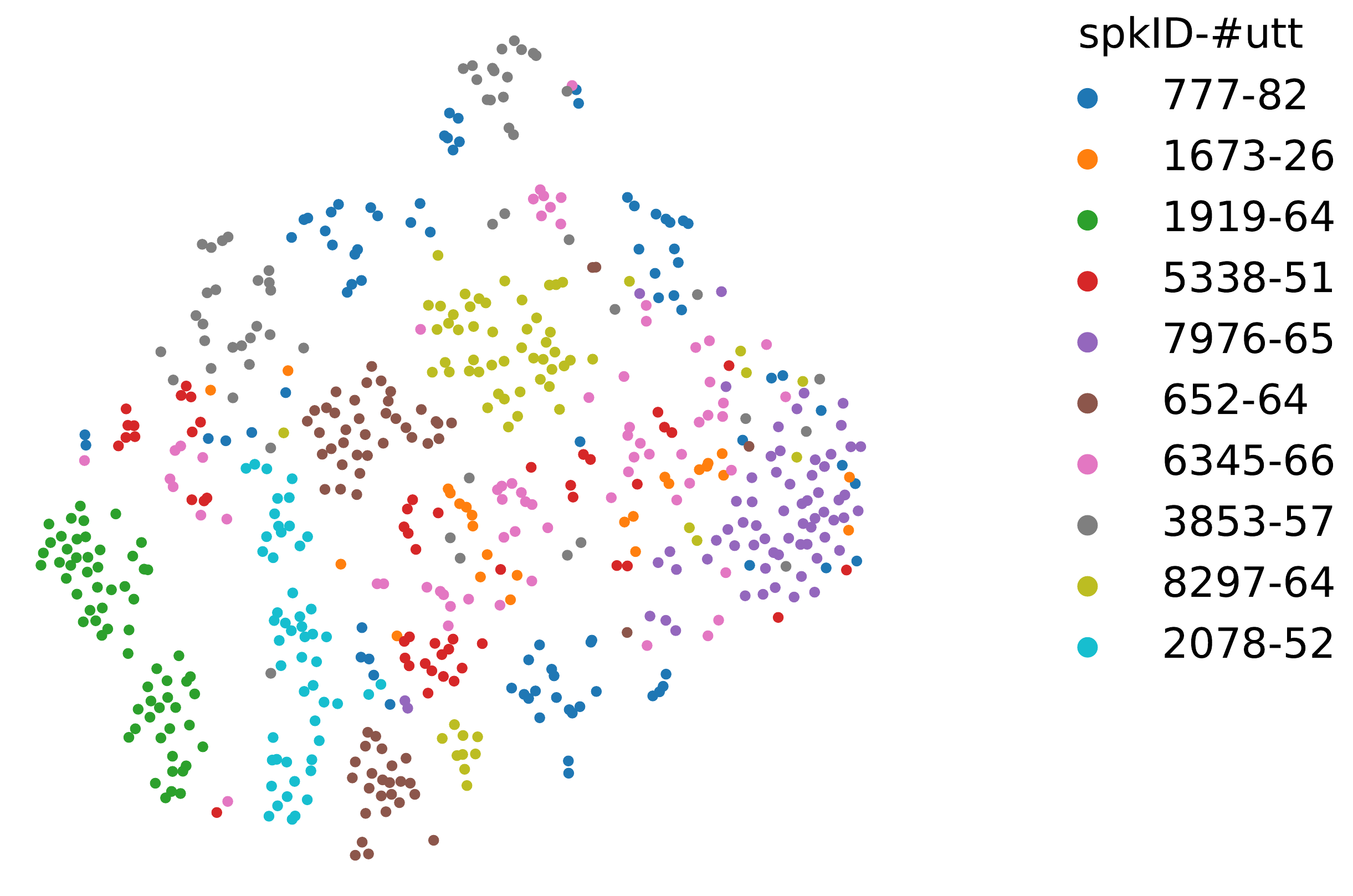}
    \caption{t-SNE visualization on libri-dev emotion embedding space}
    \label{fig:tsne}
\end{figure}

We extract emotion embeddings by models trained with IEMOCAP for 10 randomly selected speakers from libri-dev set and plot the embedding space by projecting it to 2D space using t-SNE. As observed from Figure \ref{fig:tsne}, although the embeddings are learned by training models to classify four emotions, embeddings from distinct speakers are distributed apart, while embeddings from the same speaker are clustered together in the representation space. This suggests that emotion embeddings carry a certain amount of speaker information, leading to the trade-off in speech anonymization.

To explore how much SER and ASV systems depend on the same speech characteristics for recognition, we employ the emotion recognizer as an utterance-level representation extractor and train a speaker verification model solely using the emotion representations. The emotion embeddings serve as input, followed by a hidden layer with 192 neurons, a dropout rate of 0.5, and the ReLU activation function to map the input to speaker embeddings. A speaker classifier is then employed to predict the target speaker. The training architecture and hyperparameter settings adhere to the challenge’s recipes, except that the input features are changed to emotion embeddings, and the speaker embedding extractor is simplified to a hidden layer instead of using the ECAPA-TDNN structure.

The experiment is conducted on two datasets: libri-train-360 and VoxCeleb1. For both datasets, we split the data into training and validation subsets with a ratio of 0.9 and 0.1, respectively. For the model trained on the libri-train-360 dataset, we set the epochs to 50, while for the VoxCeleb1 dataset, we set the epochs to 200. After training, we select the model with the best performance on the validation set for evaluation. For each experiment, we use 5 different random seeds, training and evaluating the model on the corresponding data.

The verification performance on the evaluation sets is shown in Table~\ref{table:emoexp}. Speakers from the test sets are distinguishable with models trained solely on emotion embeddings. Specifically, the model trained on libri-train-360 achieves an EER of 19.28\% on the lib-dev-f set and EERs of less than 10\% on other evaluation sets. The model trained with VoxCeleb1 achieves an EER of approximately 12.68\% on the corresponding test set. These results indicate that a certain level of speaker information is embedded in the emotion embeddings. Such information can be extracted and learned by a single linear layer, highlighting the challenge of disentangling speaker and emotion attributes to fully conceal speaker identity while preserving the emotional state in speech anonymization.

\begin{table}[h]
    \scriptsize
    \setlength\extrarowheight{1.4pt}
    \setlength{\tabcolsep}{4pt}
    \caption{Speaker verification results}\
    \label{table:emoexp}
    \centering
    \begin{tabular}[c]{cccccc}
        \toprule
        \multirow{2}{*}{\textbf{Train Set}} & \multirow{2}{*}{\textbf{Test Set}} & \multirow{2}{*}{\textbf{\#trials}} & \multicolumn{3}{c}{\textbf{ASV EER} (\%)} \\
        & & & mean & min & max \\
        \midrule
        \multirow{4}{*}{libri-train-360} & lib-dev-f & 15270 & 19.280 $\pm$ 3.414 & 16.620 & 23.131 \\
        & lib-dev-m & 13440 & 4.996 $\pm$ 0.583 & 4.502 & 5.433 \\
        & lib-test-f & 11744 & 9.922 $\pm$ 1.768	& 8.395 & 12.042 \\
        & lib-test-m & 9906 & 5.883 $\pm$ 0.724 & 5.120 & 6.412 \\
        \midrule
        \multirow{3}{*}{VoxCeleb1-dev} & vox1-test & 37611 & 12.686 $\pm$ 0.214 & 12.515 & 12.940 \\
         & vox1-test-f & 7036 & 16.919 $\pm$ 0.660 & 16.351 & 17.781 \\
        & vox1-test-m & 22483 & 15.621 $\pm$ 0.257 & 15.349 & 15.892 \\
        \bottomrule
    \end{tabular}
\end{table}

\section{Discussions and Conclusions}
To explore factors that expose speaker identity, we use VC approaches without paralinguistic attribute control in our study. Future work will investigate VC systems~\cite{chen2022speaker, du2021disentanglement} that incorporate speaker-emotion disentanglement abilities in the anonymization setting. The XTTS system’s ability to clone the target speaker’s voice and speaking style, along with the better emotion preservation performance of the cascaded ASR-TTS system when cloning voice-converted utterances rather than random utterances, demonstrates the effectiveness of emotion cloning. Since there is no current standard for objective evaluation of emotion cloning in the speech synthesis field, our results indicate that the emotion recognition pipeline from the VoicePrivacy 2024 challenge could be well-suited for this purpose.

While our experiments used prompt speakers and utterances from the LibriTTS dataset, using anonymized voices from more expressive corpora with richer emotions could provide a clearer insight into the entanglement between speaker information and emotion. Additionally, our study did not address the degree of speaker information exposure between speaking style and timbre, which remains unknown. The emotion recognizer trained with the IEMOCAP dataset, which has a limited number of speakers, contains retrievable speaker information. It would be interesting to see whether an emotion recognizer trained on a dataset with a larger number of speakers, like MSP-Podcast, retains more speaker information. This would be useful in understanding the speech attributes influencing data-driven speaker and emotion recognizers.

In summary, this paper explores the entanglement between speaker and emotion in speech. Our experimental results on the speech anonymization task demonstrate that enhancing privacy preservation performance results in decreased emotion preservation, highlighting the trade-off between these two attributes. However, this trade-off can be overcome if the anonymization system incorporates a robust emotion recognizer. Furthermore, emotion recognizers also retain speaker information, suggesting that speaker and emotion recognizers depend on similar speech characteristics for recognition. Thus, disentangling emotion and speaker attributes from speech remains a challenging and significant task to address.

\subsection*{Acknowledgement} 
This work was supported by the Office of the Director of National Intelligence (ODNI), Intelligence Advanced Research Projects Activity (IARPA), via the ARTS Program under contract D2023-2308110001. The views and conclusions contained herein are those of the authors and should not be interpreted as necessarily representing the official policies, either expressed or implied, of ODNI, IARPA, or the U.S. Government. The U.S. Government is authorized to reproduce and distribute reprints for governmental purposes notwithstanding any copyright annotation therein.

\newpage
\footnotesize
\bibliographystyle{IEEEtran}
\bibliography{refs}

\end{document}